\documentclass[%
 reprint,
 amsmath,amssymb,
 aps,
 pre
floatfix,
]{revtex4-1}
\renewcommand\appendix{\par 
   \setcounter{section}{0}%
   \setcounter{subsection}{0}%
   \setcounter{figure}{0}%
   \renewcommand\thesection{\Alph{section}}%
   \renewcommand\thefigure{\Alph{section}.\arabic{figure}}} 

\usepackage{graphicx}
\usepackage{dcolumn}
\usepackage{bm}

\begin{document}

\preprint{APS/123-QED}

\title{Transitions between imperfectly ordered crystalline structures: A phase switch Monte Carlo study}

\author{Dorothea Wilms$^1$, Nigel B. Wilding$^2$ and Kurt Binder$^1$}
\affiliation{$^1$Institut f\"ur Physik, Johannes Gutenberg Universit\"at Mainz,
Staudinger Weg 7, 55099 Mainz, Germany}
\affiliation{$^2$Department of Physics, University of Bath, Bath BA2 7AY, United Kingdom} %

\date{\today}

\begin{abstract}

A model for two-dimensional colloids confined laterally by ``structured
boundaries'' (i.e., ones that impose a periodicity along the slit) is
studied by Monte Carlo simulations. When the distance $D$ between the
confining walls is reduced at constant particle number from an initial
value $D_0$, for which a crystalline structure commensurate with the
imposed periodicity fits, to smaller values, a succession of phase
transitions to imperfectly ordered structures occur. These structures
have a reduced number of rows parallel to the boundaries (from $n$ to
$n-1$ to $n-2$ etc.) and are accompanied by an almost periodic strain
pattern, due to ``soliton staircases'' along the boundaries. Since
standard simulation studies of such transitions are hampered by huge
hysteresis effects, we apply the phase switch Monte Carlo method to
estimate the free energy difference between the structures as a function
of the misfit between $D$ and $D_0$, thereby locating where the
transitions occurs in equilibrium. For comparison, we also obtain this
free energy difference from a thermodynamic integration method: the
results agree, but the effort required to obtain the same accuracy as provided by
phase switch Monte Carlo would be at least three orders of magnitude
larger. We also show for a situation where several ``candidate
structures'' exist for a phase, that phase switch Monte Carlo can
clearly distinguish the metastable structures from the stable one.
Finally, applying the method in the conjugate statistical ensemble
(where the normal pressure conjugate to $D$ is taken as an independent
control variable) we show that the standard equivalence between the
conjugate ensembles of statistical mechanics is violated.

\end{abstract}

\pacs{Valid PACS appear here}
\maketitle


\section{Introduction}

Periodically ordered arrays of nanoparticles, colloidal crystals,
crystalline mesophases formed from surfactant molecules or block
copolymers, etc. are all examples of complex periodic structures that
can occur in soft matter systems. Since often the interactions between
the constituent particles of these structures are to a large degree tunable, one
has the possibility of producing materials with ``tailored'' properties
which have potential applications in nanotechnological devices
\cite{1,2,3,4,5}. When seeking to provide theoretical guidance for
understanding structure-property relations in such complex soft matter
systems, a basic issue is how to judge the relative stability of
competing candidate structures, i.e. to distinguish the stable structure
(having the lowest free energy) from the metastable ones. For standard
crystals formed from atoms or small molecules, this question can be
answered by comparing ground state energies of the competing structures
(and --if necessary-- also taking entropic contributions from lattice
vibrations into account, within the framework of the harmonic
approximation). In soft matter systems, disorder in the structure and
thermally driven entropic effects rule out such an approach, and hence
there is a need for computer simulation methods that compute the free
energy of the various complex structures. However, as is well known, the
free energy of a model system is not a direct output of either
Molecular Dynamics or Monte Carlo simulations, and special techniques
have to be used \cite{6,7,8,9,10,11}.

In principle, one can obtain the absolute free energy of a structure by
linking it to some reference state of known free energy by means of
thermodynamic integration (TI)
\cite{6,7,8,9,10,11,12,13,14,15,16}. The strengths of TI are that
it is both conceptually simple and often straightforward to implement.
Its principal drawback is that the quantity of interest, namely the free
energy {\em difference} between candidate structures is typically
orders of magnitude smaller than the absolute free energies of the
individual structures which TI measures. Essentially, therefore, TI
estimates a small number by taking the difference of two large ones; As
a consequence, the precision of the method is limited and an enormous
(even sometimes wasteful) investment of computer resources may be needed
to resolve free energy difference accurately \cite{9}.

A much more elegant approach, albeit one which is not quite so easy to
implement as TI, is the ``phase switch Monte Carlo'' \cite{17a,17,18,19,20,21,21a}
technique. This method is potentially more powerful than TI because it
focuses directly on the small free energy difference between the
structures to be compared, rather than their absolute free energies. In
previous work, the precision of the method was demonstrated in the
context of measurements of the free energy difference between fcc and
hcp structures of hard spheres \cite{19}, the phase behaviour of
Lennard-Jones crystals \cite{19} and as a means of studying liquid-solid
phase transitions \cite{17}. In the latter case, simple model systems
containing only a few hundred particles could be studied, while for the
study of the fcc-hcp free energy difference \cite{17a,20} larger systems
of up to $1728$ particles could be studied by virtue of the fact that
these crystals only differ in their packing sequence of close-packed
triangular defect-free lattice planes. However, it is an open question
as to what system sizes one can attain with the phase switch method for
more general crystalline systems, including -- as in the present work --
ones which exhibit considerable structural disorder (``soliton
staircases'', see below). Furthermore, there have hitherto been no
like-for-like comparisons of the TI and phase switch methods on the same
system, so whilst their are good reasons for {\em presuming} the
superiority of phase switch (in terms of precision delivered for a given
computational investment), this has never actually been quantified.

In the present paper, we address these matters, considering as a generic
example a two-dimensional colloidal crystal in varying geometrical
confinement \cite{22,23,24,25,26}. As is well-known, two dimensional
colloidal crystals are experimentally much studied model systems
\cite{27,28,29,30,31,32,33,34,35,36,37,38} comprising, for example,
polystyrene spheres containing a superparamagnetic core adsorbed at the
air-water-interface. Applying a magnetic field oriented perpendicular to
this interface creates a repulsive interaction that scales like $r
^{-3}$, ($r$ being the particle separation), whose magnitude is
controlled by the magnetic field strength \cite{27}. Lateral
confinement of such two-dimensional crystals can be effected
mechanically or by laser fields (if the latter are also applied in the
bulk of such a crystal, one can study laser-induced melting and/or
freezing \cite{39,40,41,42}). Of course, there exist many related
problems in rather different physical contexts (``dusty plasmas''
\cite{43,44}, i.e. negatively charged SiO$_2$ fine particles with 10$\mu
m$ diameter in weakly ionized $rf$ discharges; lattices of confined
spherical block copolymer micelles \cite{45}; vortex matter in slit
channels \cite{46}, etc.). However, our study does not address a
specific system, rather we focus on the methodological aspects of
how one can study such problems by computer simulation. 

The outline of the present paper is a follows. In Sec. 2, we summarize
the key facts about our model, namely strips of two-dimensional crystals
confined between two walls where structural phase transitions may occur
when the distance between the (corrugated) rigid boundaries is varied
\cite{23,24,25,26,47,48,49} (i.e., a succession of transitions in the
number of crystal rows $n$ parallel to the walls occur, $n \rightarrow
n-1 \rightarrow n-2$, with increasing compression, accompanied by the
formation of a ``soliton staircase'' at the walls \cite{23,24,25,26}). In
Sec. 3, the methods that are used are briefly described: the
thermodynamic integration method of Schmid and Schilling \cite{15,16} is
used as a baseline, while the main emphasis is on the phase switch Monte
Carlo method (implementation details of which are deferred to an
Appendix). In Sec. 4 we describe the results of the application of these
techniques to the model of Sec. 2. We show that phase switch Monte Carlo
\cite{17,18,19} can accurately locate the phase transitions despite the
need to deal with thousands of particles, and is orders of magnitude
more efficient than thermodynamic integration. Sec. 5 summarizes some
conclusions.

\section{Structural Transitions in Crystalline Strips confined by corrugated boundaries: Phenomenology}

Here we introduce the model for which our methodology is exemplified,
and recall briefly the main findings concerning the rather unconventional
transitions that have been detected \cite{23,24,25,26}, as far as they
are relevant for the present study.

We consider monodisperse colloidal particles in a strictly
two-dimensional geometry, which then are treated like point particles in
a plane interacting with a suitable effective potential $V(r)$ that
depends only on the interparticle distance $r$. In the real systems
\cite{27,29,30,31,32,33} this potential is purely repulsive, but due to
the magnetostatic dipole-dipole interaction (whose strength is
controlled by the external magnetic field) it is very slowly decaying,
$V(r) \propto r ^{-3}$. Since we here are not concerned with
quantitative comparisons with real experimental data on such systems, we
simplify the problem by adopting a computationally more efficient
$r^{-12}$ potential, in accord with previous work \cite{23,24,25,26}.
Moreover, to render it strictly short-ranged, we introduce a cutoff
$r_c$, such that $V(r \geq r_c)\equiv 0$, and employ a smoothing function
to make $V(r)$ differentiable at $r=r_c$. Thus, the model potential used
is

\begin{equation} \label{eq1}
V(r) = \varepsilon\Big[(\sigma/r)^{12} - (\sigma/r_c)^{12} \Big] \Big[\frac{(r-r_c)^4}{h^4 + (r-r_c)^4}\Big] \quad ,
\end{equation}
with parameters $r_c=2.5\sigma$ and $h=0.01 \sigma$. Henceforth, the
particle diameter $\sigma=1$ defines the length units in our model, and
for the energy scale, $\varepsilon=1$ is taken, while Boltzmann's constant
$k_B=1$. It is known that at $T=0$ the ground state of this model is a
perfect triangular lattice, with a lattice spacing $a$ related to
the choice of number density $\rho=N/V$ (with $N$ the particle number and $V$ the
(two-dimensional) ``volume'' of the system) via

\begin{equation} \label{eq2}
a^2 =2 / (\sqrt{3} \rho) \quad .
\end{equation}

Assuming the physical effect of truncating the potential can be
neglected, only the choice of the combination $X=\rho
(\varepsilon/k_BT)^{1/6}$ controls the phase behavior \cite{49a}.
Thus, following previous work in the NVT-ensemble it
suffices to choose a single density when the temperature variation is
considered \cite{23,50}. For the particular choice $\rho=1.05$, the melting
transition of this model is known to occur at about $T=T_m \approx 1.35$
\cite{50}. Note that here we are not at all concerned with the
peculiarities of melting in two dimensions \cite{51}, and hence we focus
on a temperature deep within the crystalline phase, $T=1$. Although it
is known that the density of vacancies and interstitials in $d=2$ for
any nonzero temperature is also nonzero in thermal equilibrium
\cite{51,52}, for the chosen particle number $N= 3240$ the system is
essentially defect free, since the densities of these point defects at
$T=1$ are extremely small \cite{23,50}.

\begin{figure}
\includegraphics[scale=0.28, clip=true]{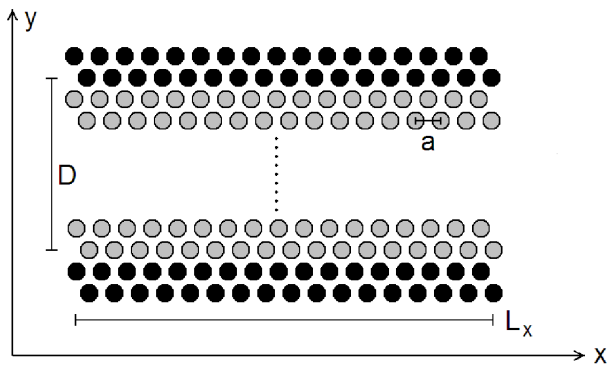}

\caption{\label{fig1} Sketch of the system geometry, showing the fixed
wall particles (black spheres) and the mobile particles (gray spheres).
The orientation of the coordinate axes is indicated, as well as the
lattice spacing of the triangular lattice ($a$) and the linear dimensions
$L_x,D$ of the system.}

\end{figure}

The geometry of the present system is a $D \times L_x$ slit, confined in
the y-direction and periodic in the x-direction. In the y-direction
there are $n_y=30$ rows of the triangular lattice, each containing
$n_x=108$ particles, stacked upon each other. The $x$-direction
coincides with a lattice direction so that $L_x=n_xa$. The
confining boundaries (one at the top and one at the bottom of the
system) each take the form of a double rows of particles in which the
particles are rigidly fixed at the sites of a perfect triangular lattice
(Fig.~\ref{fig1}). These rows of fixed particles represent rigid
corrugated walls, essentially acting as a periodic wall potential on the
mobile particles. While the distance of the first row at the upper wall
from the first row of mobile particles in the ideal stress-free crystal is
simply $D=n_y a \sqrt{3}/2$, in the following we are interested in the response of the
system when the walls occur at a smaller distance, caused by a misfit
$\Delta$, defined via \cite{53}

\begin{equation} \label{eq3}
D=(n_y - \Delta) a \sqrt{3} / 2 \quad .
\end{equation}

\begin{figure}
\includegraphics[scale=0.32, clip=true]{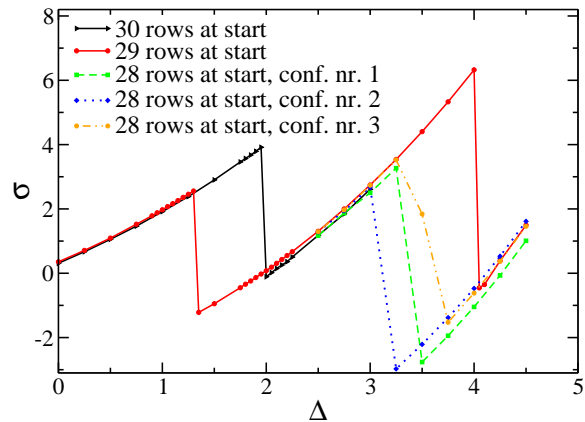}
\caption{\label{fig2} Stress $\sigma$ plotted vs. misfit $\Delta$, for a
system of $N=3240$ particles, and using different starting
configurations having $n_y=30$, $n_y=29$, and $n_y=28$, as indicated in
the figure. Note the huge hysteresis of the $n_y=30 \rightarrow n_y=29$
and $n_y=29 \rightarrow n_y=28$ transitions. For further explanations
see the main text.}
\end{figure}

\begin{figure}
\includegraphics[scale=0.1, clip=true]{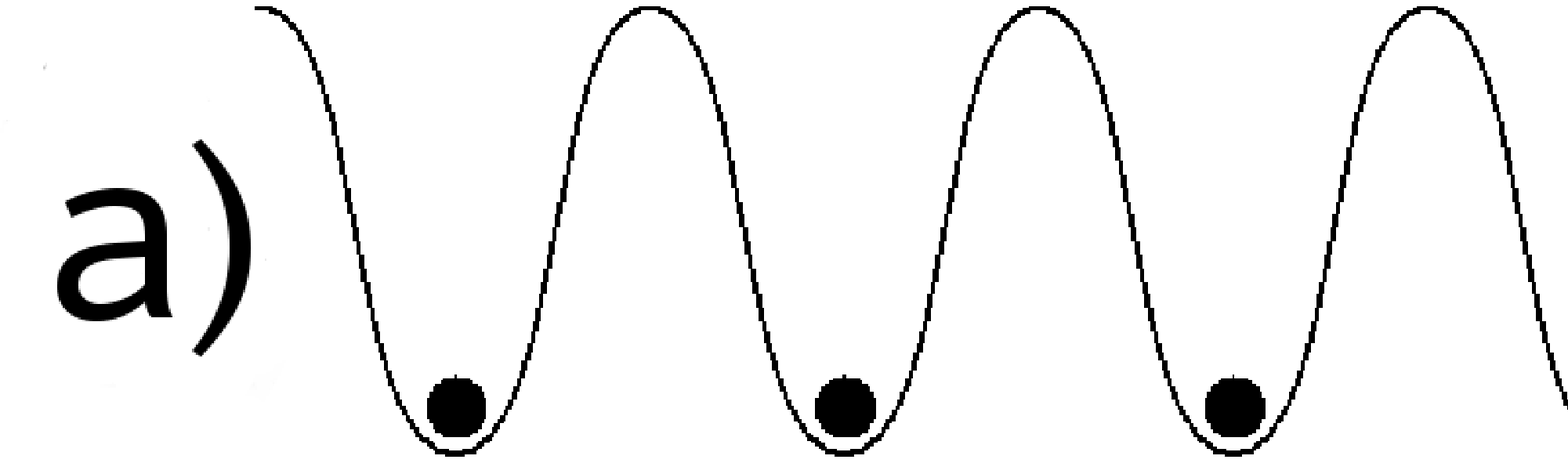}\\
\includegraphics[scale=0.18, clip=true]{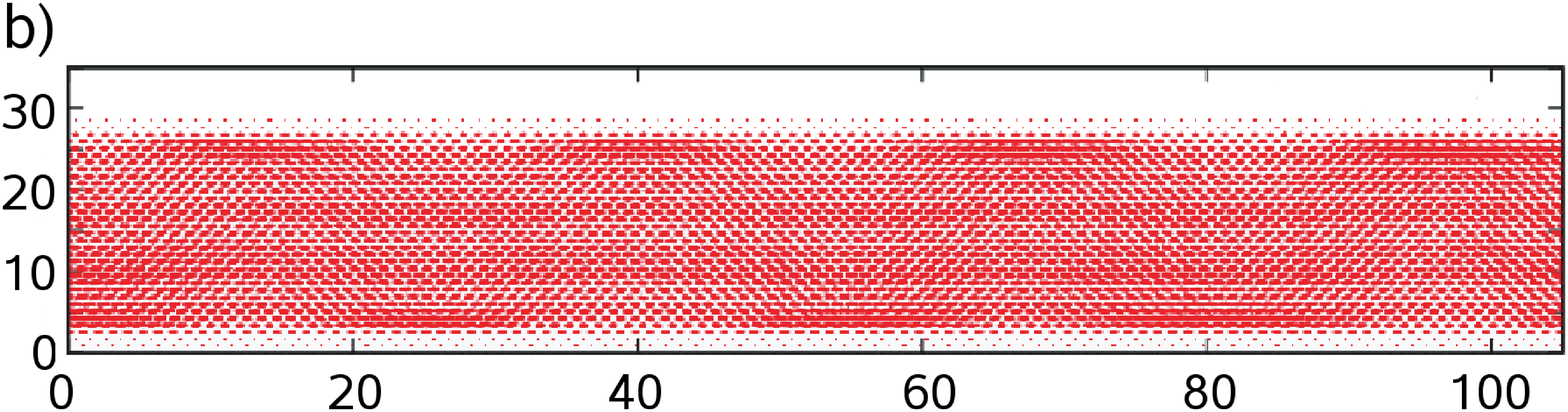}\\
\includegraphics[scale=0.18, clip=true]{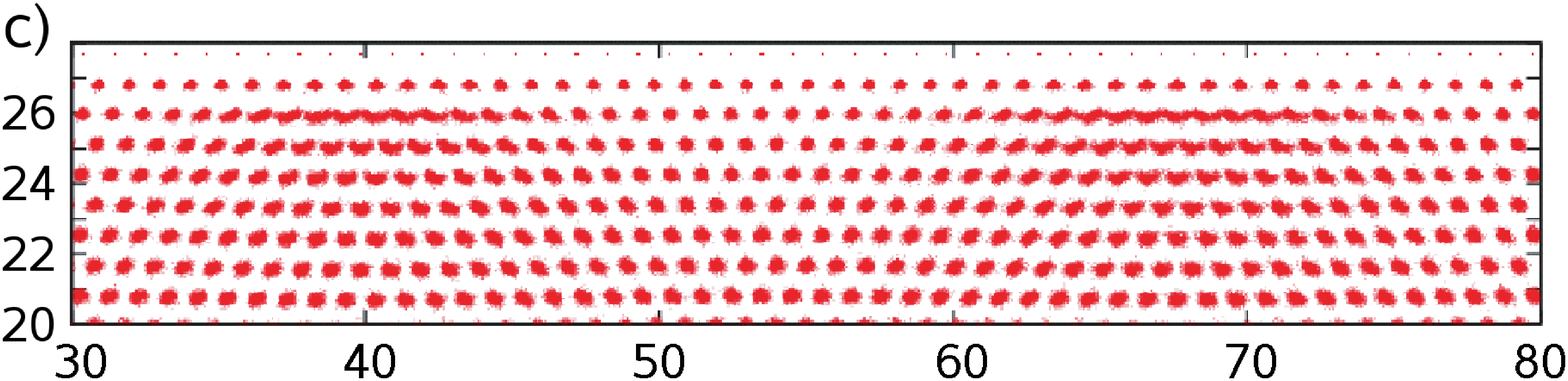}
\caption{\label{fig3} a) Putting $n+1$ particles in a periodic potential with $n$ minima creates a soliton configuration,
i.e. over a range of several lattice spacings particles are displaced from the potential minima (schematic) b) Superimposed snapshot pictures of 750 configurations of the particle positions, where for a system of $n_y=30$ rows and a large misfit ($\Delta=2.6$) a transition to $n_y-1=29$ rows has occured ($n_x=108$ and $T=1.0$ were chosen). The $4$ solitons at each wall are visible due to the larger lateral displacements of the particles, leading to a darker region in the snapshot. Part (c) shows a close-up of the structure near the upper wall. Numbers shown along the axes indicate the Cartesian coordinates of the particles. Parts (b) and (c) have been adapted from Chui et al. \cite{23}.}
\end{figure}

As described in the previous work \cite{23,24,25,26}, standard
Monte Carlo simulation \cite{6,7} allows one to study this model at
various values of $\Delta$, and also sample the stress
$\sigma=\sigma_{yy} - \sigma_{xx}$ ($\sigma_{\alpha \beta}$ are the
Cartesian components of the pressure tensor) applying the virial formula
\cite{6,7}. Fig.~\ref{fig2} shows that when one starts out with the
perfect crystal $(n_y=30)$ with no misfit, the crystal already shows a
small finite stress, because the rigid wall particles somewhat hinder
the vibrations of the mobile particles in their potential wells, but
this effect is of no importance here. Rather we focus on the (slightly
nonlinear) increase of the stress up to about $\Delta =\Delta_c \approx
2$, followed by the (almost) discontinuous decrease, and the subsequent
increases again with further enhancement of the misfit. A previous
structural analysis has revealed \cite{23,24,25,26} that the sudden
decrease of stress is due to a transition in the number of rows in the
crystal, $n_y \rightarrow n_y -1=29$. However, since in the NVT ensemble
the particle number is conserved, the $n_x=108$ particles of the row
that disappears must be redistributed among the remaining rows. A
closer examination of the structure revealed that none of these
particles enter the two rows adjacent to the rigid walls, instead they
all go into the $n_y-3=27$ rows of the system that are further away from the
walls. Thus, in the present case, the particle number per row becomes
$n'_x+n_x/(n_y-3)=n_x+4$, and this leads to a new lattice spacing in the
$x$-direction of $a'=a/(1+4/n_x)$, which is no longer commensurate with
the spacing between the particles forming the rigid walls (or the two immediately adjacent
layers which remain commensurate with them). While for the rows in the
center of the system (near $n_y/2)$ this compression of the lattice
spacing occurs uniformly along the $x$-direction, this is not the case
close to the walls, which provide a periodic potential (with periodicity
$a$) that acts on the row of mobile particles a little further inside
the slit. The fact that on the scale $L_x$ the effective wall potential
exhibits $n_x$ minima but that  $n'_x=n_x+4$ particles need to be
accommodated, leads to the formation of a lattice of solitons close to
both walls (``soliton staircase'') \cite{54,55}, as depicted for an
idealized case in Fig.~\ref{fig3}.

In practice,  the actual structure having $n_y-1=29$ rows that is formed
in the simulations on increasing the misfit $\Delta$ beyond the critical
value $\Delta_c$, is generally less regular than the 'idealized' one
shown in Fig.~\ref{fig3}: specifically, the relative distance between
neighboring solitons showed a considerable variation. This comes about
because (i) the solitons are formed from the stressed crystal with
$n_y=30$ rows via random defect nucleation events \cite{24}, and (ii)
the mutual interaction between neighboring solitons, which is the
thermodynamic driving force towards a regular soliton arrangement, is
very small \cite{25}. Despite this, it is nevertheless reasonable to
construct ``by hand'' the expected regular structure of $n_x/(n_y-3) \,
(=4)$ solitons near each wall as a starting configuration for a system
with $29$ rows, which can subsequently be equilibrated \cite{23}. Of
course, there is no guarantee that this guessed structure actually is
the one of lowest free energy; but it does exhibit slightly less stress
than all other structures that had been tested, for misfits in the range
$1.5 \leq \Delta \leq 3$, and hence has been used as a starting point
for studies in which $\Delta$ was varied in this range. 

Starting from this idealized $29$ row structure and decreasing the
misfit one finds that the $29$ row structure is stable down to about
$\Delta'_c\approx 1.3$, at which point the soliton lattice disappears
and the system spontaneously transforms into a defect free structure
with $n_y=30$ rows again (Fig.~\ref{fig2}). This value of $\Delta$ is to
be compared with that for the reverse transition from $30$ to $29$ rows
which we recall occurs at $\Delta_c \approx 2.0$. Thus, with the standard
Monte Carlo approach there is considerable hysteresis which precludes
the accurate location of the transition point.  Clearly, therefore a
method is needed from which one can locate where the transition occurs
in equilibrium. 

\begin{figure}
\includegraphics[scale=0.4, clip=true]{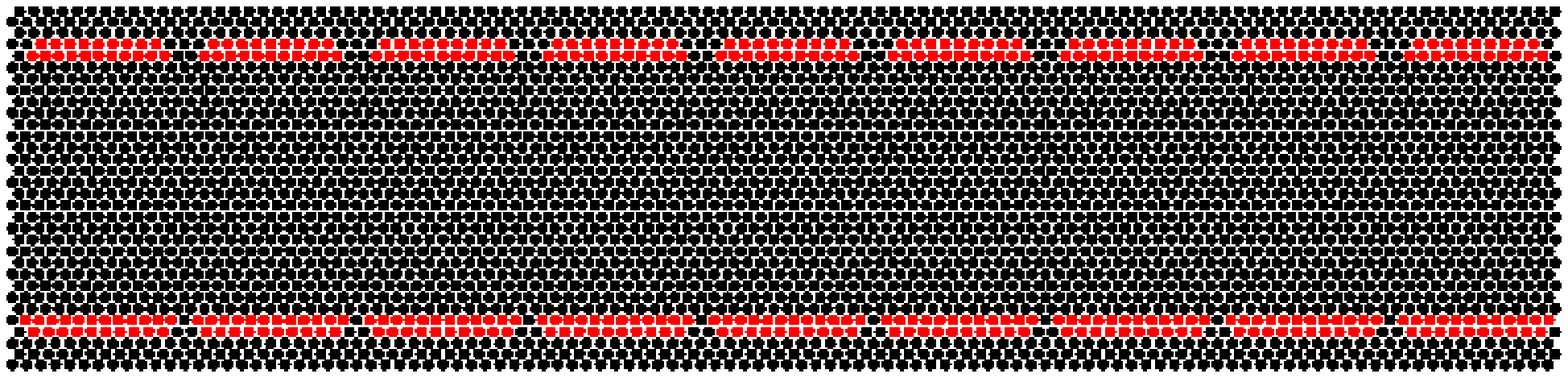}\\
\includegraphics[scale=0.4, clip=true]{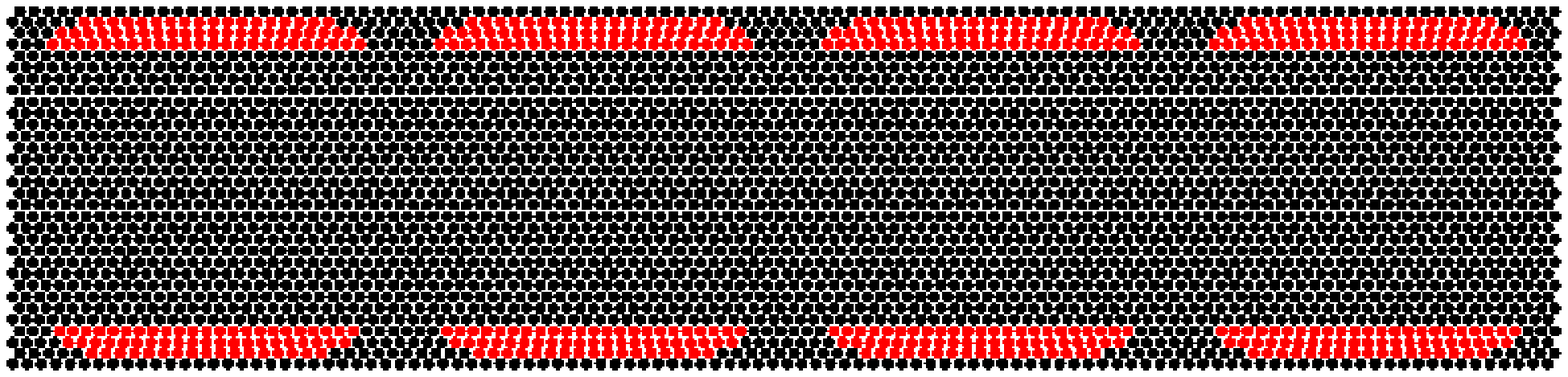}\\
\includegraphics[scale=0.4, clip=true]{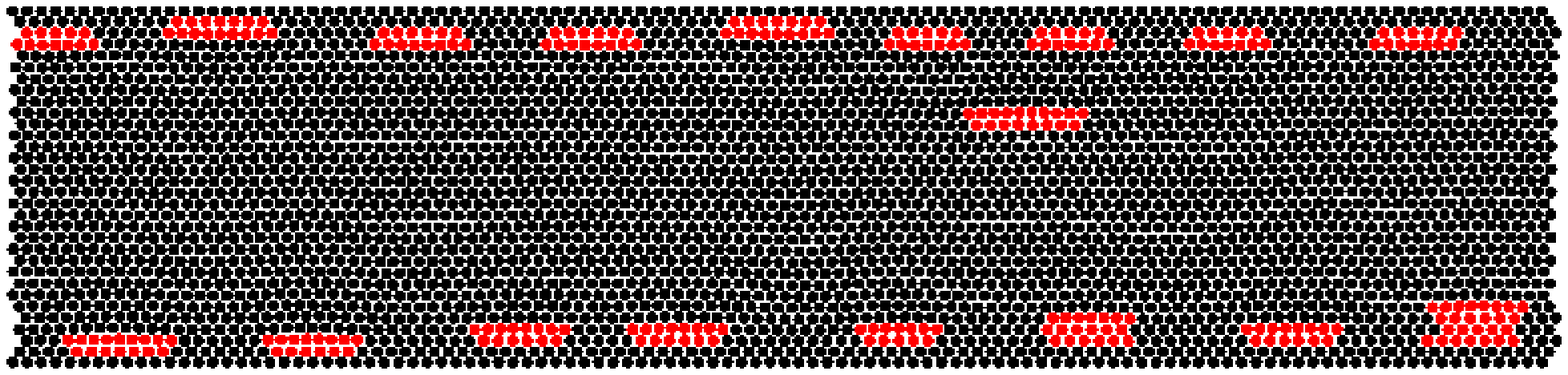}\\
\includegraphics[scale=0.4, clip=true]{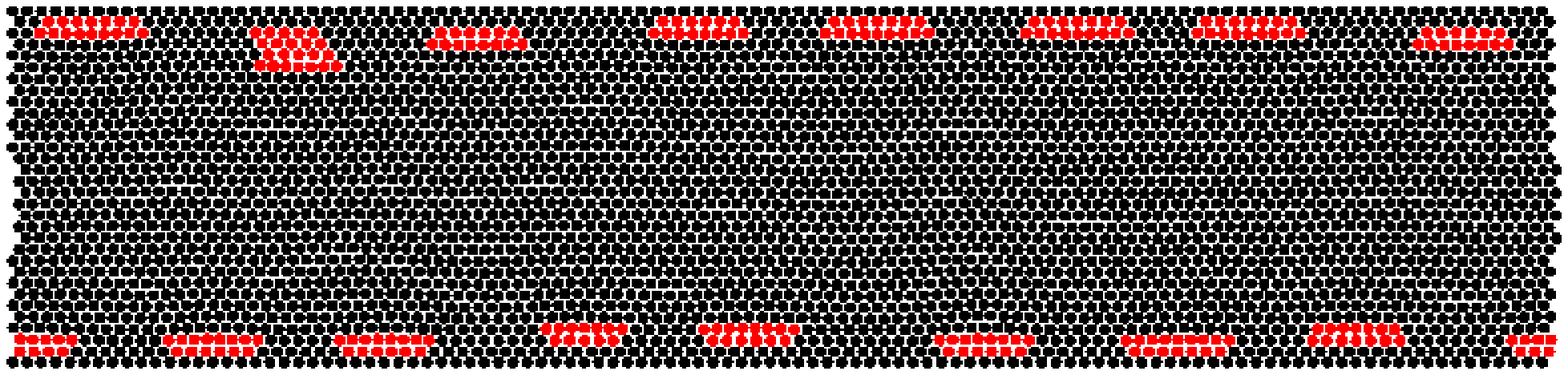}\\
\caption{\label{fig4} Configurations with $N=3240$ particles and $n_y-2=28$ rows, but different configurations of the solitons. In the text, they are referenced as ``configuration nr.~1,~2,~3,~4'' from top to bottom. For a clear identification of the positions of the solitons, the method described in \cite{25} was used.}
\end{figure}

Similar hysteresis is observed if one starts out from the $29$ row
structure but increases the misfit beyond $\Delta =3$ (a case that has
not been studied previously). As Fig.~\ref{fig2} shows, a transition occurs
to structures with $n_y-2=28$ rows (at about $\Delta \approx 4.1$).
Unfortunately, there seem to be no unique candidates for stable
structures having $n_y-2=28$. Fig.~\ref{fig4} displays four candidate
structures that we have identified, each of which is at least metastable
on simulation timescales. Depending on which of these $28$ row
candidates one takes, the transition from $28$ to $29$ rows on reducing
the misfit occurs at anything between $\Delta=3.2$ and $3.75$. As
regards the nature of the candidate structures, in each case $2n_x=216$
extra particles have to be distributed across the system. If we again
keep the rows adjacent to the walls free of extra particles, the
particle number per inner row becomes $n'_x=n_x + 2 n_x/(n_y-4)\approx
n_x + 8.3$, i.e. is non-integer. If we kept two rows adjacent to the wall
rows free of extra particles, we would have $9$ extra particles per row,
and thus this structure has been tried (this is configuration number $1$
in Fig.~\ref{fig4}). Another structure was obtained if we place $4$ extra
particles in the rows directly adjacent to the walls and $8$ extra
particles in each of the $26$ inner rows (configuration number $2$). By
energy minimization of a somewhat disordered structure resulting from a
transition from $29$ to $28$ rows a structure was obtained which had $9$
solitons on one wall but only $8$ on the other wall (configuration
number $3$). Finally another configuration with $8$ solitons on each
wall (configuration number $4$) was found. Note that the configurations shown in Fig.~\ref{fig4}
are not the actual structures at $T=1.0$ but the corresponding ``inherent structures''
found from the actual structures by cooling to $T=0$, to clearly display where the solitons occur.
Clearly, it again is a
problem to (i) identify which of these $4$ configurations with $28$ rows
is the stable one (at $T=1.0$), and (ii) determine at which misfit the transition to the
structure with $29$ rows occurs. As we shall demonstrate below, both
problems can be elegantly dealt with by employing the phase switch Monte Carlo
method.

\section{Free energy based simulation methodologies to locate
transitions between imperfectly ordered crystal structures}

\subsection{Thermodynamic Integration}

The general strategy of thermodynamic integration is to  consider a
Hamiltonian $\mathcal{H} (\lambda)$ that depends on a parameter
$\lambda$ that can be varied from a reference state (characterized by
$\lambda_0$) whose free energy is known, to the state of interest
$(\lambda_1)$, without encountering phase transitions. The free energy
difference $\Delta F$ can then be written as

\begin{equation} \label{eq4}
\Delta F= F (\lambda_1) - F (\lambda_0) = \int\limits_{\lambda_0}^{\lambda_1} d \lambda' \langle \partial \mathcal{H}(\lambda')
/\partial \lambda' \rangle_{\lambda'} \quad .
\end{equation}

For a dense disordered system (fluid or a solid containing defects),
Schilling and Schmid \cite{15,16} proposed to take as a reference state a
configuration chosen at random from a well equilibrated simulation of
the structure of interest, at values of the external control parameters
for which one wishes to determine the free energy. Particles can be
held rigidly in the reference configuration $\{\vec{r}_i \,^ {\rm ref}\}$ 
by means of a suitable external potentials. (We recall that a somewhat related
thermodynamic integration scheme for disordered systems known as the
``Tethered spheres method'' has already been proposed by Speedy
\cite{55a}.) When these external potentials act, the internal
interactions can be switched off. In practice, one can use the following
pinning potential $U_{\rm ref} (\lambda)$ to create the reference state,
where $r_{\rm cut}$ is a parameter discussed below.

\begin{equation} \label{eq5}
U_{\rm ref} (\lambda)= \lambda \sum\limits_i \phi (|\vec{r}_i - \vec{r}\;^{\rm ref}_i |/r_{\rm cut}) \quad {\rm with}\,
\phi\, (x)=x-1 \quad .
\end{equation}

Here it is to be understood that particle $i$ is only pinned by well $i$ at
$\vec{r}\;^{\rm ref}_{i}$, and not by other wells. However, identity swaps
need to be carried out to ensure the indistinguishability of particles.
The free energy of this non-interacting reference system then is

\begin{equation} \label{eq6}
F_{\rm ref} (\lambda) =\ln (N/V) -\ln [1+ (V_0/V) g_\phi (\beta\lambda)]\:,
\end{equation}
where $\beta=(k_B T)^{-1}$, $V_0$ (in $d=2$ dimensions) is $V_0= \pi r^2 _{\rm cut} $ and

\begin{eqnarray} \label{eq7}
&& g_\phi (a) = \frac{2}{\lambda^2} [\exp (a) - \sum\limits_{k=0}^2 e^k / k!]\:,\nonumber\\
\end{eqnarray}
for the choice of $\phi(x)$ written in Eq.~(\ref{eq5}).

Then intermediate models $\mathcal{H}(\lambda)$ to be used in Eq.~(\ref{eq4}) are chosen as

\begin{equation} \label{eq8}
\mathcal{H}' (\lambda) = \mathcal{H}_{\rm int} + U_{\rm ref} (\lambda) \quad ,
\end{equation}
where $\mathcal{H}_{\rm int}$ describes interactions in the system,
which then are switched on (if necessary, in several steps). The free
energy contribution of switching on these interactions can easily be
determined by a Monte Carlo simulation which includes a move that
switches the interactions on and off. The logarithm of the ratio of how
many times the states with and without interactions were visited gives
the free energy contribution. The free energy difference between the
intermediate model where particle interactions are turned on and
potential wells are also turned on, and the target system with particle
interactions but without potential wells, then is computed by
thermodynamic integration, for which

\begin{equation} \label{eq9}
\langle \partial \mathcal{H}_{\rm ref} (\lambda) / \partial \lambda \rangle = \langle \sum_i \phi (
|\vec{r}_i - \vec{r}_i\;^{\rm ref}|/ r_{\rm cut}) \rangle
\end{equation}
needs to be sampled \cite{15,16}. This method has been tested for hard spheres \cite{15,16},
including also systems confined by walls from which wall excess free energies could be
sampled \cite{56}.

\subsection{Phase Switch Monte Carlo}

The phase switch method \cite{17,18,19,20,21,21a} computes directly the
relative probabilities of two phases, by switching between them and
recording the ratio of the simulation time spent in each. This ratio
directly yields their free energy difference $\Delta F$ via $\Delta F=
\ln(A^{(1)}/A^{(2)})$. Here  $A^{(1)}$ and $A^{(2)}$ are the times spent
in the respective phases which are proportional to the statistical
weight of each phase \cite{9}.

\begin{figure}
\includegraphics[scale=0.32, clip=true]{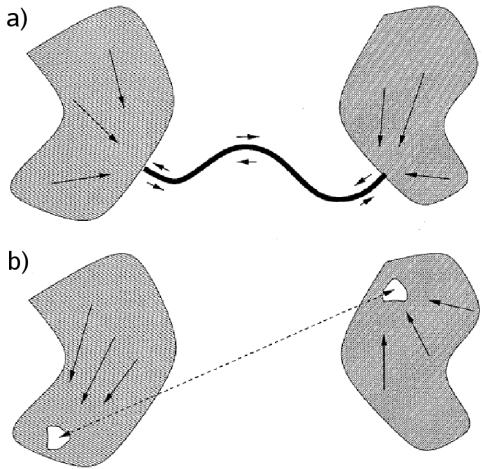}
\caption{\label{fig5} Schematic comparison of (a) the standard method
for linking phases via a sampling path and (b) The phase switch method.
The blobs represent the set of values of some macroscopic property (eg
order parameter or energy) associated with configurations belonging to
two distinct phases $(\alpha=1,2)$. These pure phase states (having high
probability) are separated by a ``deep valley'' in the free energy
landscape corresponding to interfacial states having a very low
probability. (a) In the standard strategy one uses extended sampling to
negotiate the valley, by climbing down into it from one side and climbing
up out of it on the other. (b) The idea of phase switch Monte Carlo is
to ``jump over the valley''.}
\end{figure}

The power of the phase switch method derives from its ability to leap
directly from configurations of one pure phase to those of another pure
phase (Fig.~\ref{fig5}), avoiding the mixed phase states which -- when
one or both phases are crystalline -- can be computationally problematic
(see appendix A). The leap is implemented as a suitable global Monte
Carlo move. One starts out by specifying for each of the two phases of
interest (labeled by index $\alpha=1,2$), a reference configuration. This
can be expressed as a set of  $i=1\ldots N$ particle positions $\{
\vec{R}_i^{\,(\alpha)}\}$. Note that the specific choice of a reference
configuration for phase $\alpha$ does not matter (at least in principle, see
Appendix), it need only be a member of the set of pure phase
configurations that ``belong'' to phase $\alpha$. Thus for example in the
present case, a suitable reference configuration for the $n=30$ row
defect-free structure could simply be a typical configuration chosen
from a simulation run on this structure. However, it could equally be a
configuration in which all particles are at the lattice sites of this
structure.

Given the two reference configurations, one can express the
position vectors $\vec{r}_i^{\,(\alpha)}$ of each particle $i$ in phase $\alpha$ as 

\begin{equation}
\vec{r}_i^{\,(\alpha)}= \vec{R}_i^{(\alpha)} + \vec{u}_i\:.
\end{equation}

where $\{\vec{u}_i\}$ is a set of displacement vectors which measure the
deviation of each particle from the reference site to which it is
nominally associated. Note that while there is a separate reference
configuration for each phase, the single set of displacements is
common to both phases. 

Let us suppose the simulation is currently in phase $\alpha=1$. Now the phase
switch idea is to a map the current configuration $\{\vec{r}_i^{\,(1)}\}$ of this
phase on to a configuration of  phase $\alpha=2$ by switching the sets of
reference sites from $\{\vec{R}_i^{\,(1)}\}$ to $\{\vec{R}_i^{\,(2)}\}$ but
keeping the set of displacements $\{\vec{u}_i\}$ {\em fixed}. This
switch can be incorporated in a global Monte Carlo move. Of
course, in general the set displacements that are typical for phase
$\alpha=1$ will not be typical displacements for phase $\alpha=2$.
As a consequence, in a naive implementation such a global move will
almost always be rejected by the Monte Carlo lottery. This problem is
circumvented by employing extended sampling methods \cite{9,10,56a} that
create a bias which enhances the occurrence of displacements
$\{\vec{u}_i\}$ for which the switch operation does have a sufficiently
high Monte Carlo acceptance probability. Such states are called
``gateway states'' \cite{17,18,19,20,21}: crucially, they do not need to
be specified beforehand - the system autonomously guides itself to them
in the course of the biased sampling.

In practice, the bias is administered with respect to an ``order
parameter'' $M$ whose instantaneous value is closely related to the
energy cost of implementing the phase switch. One then introduces a
weight function $\eta(M)$ into the sampling of the effective Hamiltonian
which enhances the probability of the system sampling configurations for
which the energy cost of the phase switch is low, thereby increasing the
switch acceptance rate. Of course, the weight function $\eta(M)$ to be
used is not known beforehand, and thus needs to be iteratively
constructed in the course of the Monte Carlo sampling. One has a choice
of ways of doing so: we have used the transition matrix Monte Carlo
method \cite{56a,57,58} (see also the Appendix for implementation
details). Alternative methods such as Wang-Landau sampling \cite{59} or
successive umbrella sampling \cite{73} could also be applied. 

Once a suitable form for the weight function $\eta(M)$ has been found, a
long Monte Carlo run is performed, in the course of which both phases
are visited many times. The statistics of the switching between phases
is monitored by accumulating the histogram of $M$, which (as in all
extended sampling methods) is corrected for the imposed bias at the end
of the simulation. Doing so yields an estimate of the true equilibrium
distribution $P(M)$, which in general exhibits a double peaked form (one
peak for each phase). The free energy difference between the two phases
is simply the logarithm of the ratio of the peak weights as described at
the start of this subsection.

Of course, the above description was only intended to outline the 
phase switch strategy; more extensive implementation details are given
in the appendix.

\section{Results}

\subsection{Free energy differences and computational efficiency}

\begin{figure}
\includegraphics[scale=0.32, clip=true]{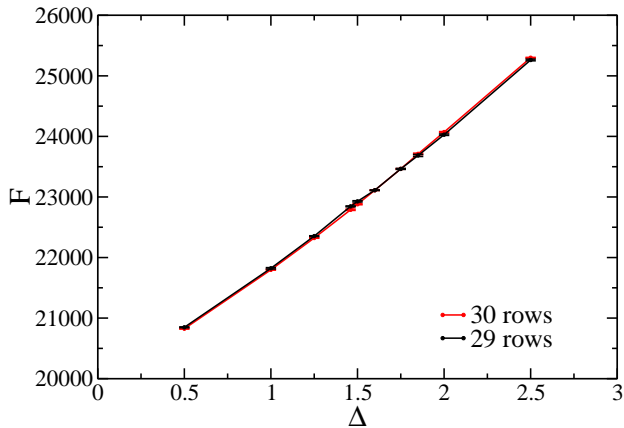}
\caption{\label{fig6}Absolute free energy $F$ of systems of $N=3240$
particles interacting with the potential given in Eq.~(\ref{eq1}) in $L
\times D$ geometry with $L=108 a$, $a$ being the lattice spacing, and
periodic boundaries in $x$-direction, confined by two rows of fixed
particles on either side in $y$-direction (Fig.~\ref{fig1}, as a
function of the misfit $\Delta$ \(Eq.~(\ref{eq3})\). Two structures are
compared:(i) a (compressed) triangular lattice with $n_y=30$ rows
containing $n_x=108$ particles per row; (ii) a lattice with $n_y=29$
rows and corresponding soliton staircases (Fig.~\ref{fig3}b).}
\end{figure}

Fig.~\ref{fig6} shows the absolute free energies in the NVT ensemble for
the phase with 30 rows (and no defects) and the phase with 29 rows and
the ``soliton staircases'' (Fig.~\ref{fig3}b) as a function of the
misfit $\Delta$, as obtained from the thermodynamic integration method
(Sec. III.1). One sees that these free energies are very large (note the
ordinate scale) and vary rather strongly with $\Delta$. However, the
free energy curves with these two structures are barely distinct from
each other, and hence a very substantial computational effort is needed
to locate, with meaningful accuracy, the intersection point marking the
equilibrium transition between $n=30$ and $n=29$ rows.

\begin{figure}
\includegraphics[scale=0.32, clip=true]{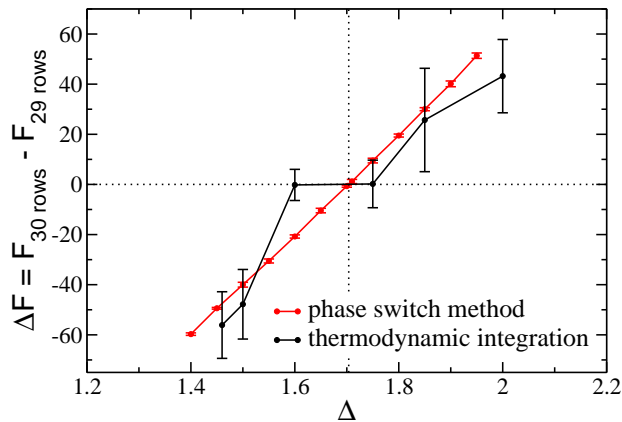}
\caption{\label{fig7} Free energy differences between structures with 29 and 30 rows plotted versus the misfit $\Delta$. Both results obtained from thermodynamic integration and from the phase switch method are shown, as indicated.}
\end{figure}

Fig.~\ref{fig7} plots the free energy difference $\Delta F$ versus the
misfit, comparing the results from the thermodynamic integration method
(points with error bars) with the results from the phase switch method,
and focusing on the region near the transition. One can see that within
the errors the results of both methods agree very well with each other,
although for the thermodynamic integration method the error is at least
an order of magnitude larger than that of phase switch. We note that the
predicted equilibrium value of the misfit at the transition point ($\Delta_t
\approx 1.7)$ falls well within the hysteresis loop of Fig.~\ref{fig2}.

Since the absolute free energies are of the order of 20000 (for our
system with $N=3240$ particles) but, in the region of interest, free
energy differences are of order $\pm 40$ only, we have that the relative
error $\delta F/F$ is of order $1/500$. Thus for thermodynamic
integration, it would be difficult to bring the error bars down further
in Fig.~\ref{fig7}. The error bars for the phase switch simulation were
computed from the results of four independent runs for each value of the
misfit, and are hardly visible on the scale of Fig.~\ref{fig7}.

In addition to this significant difference with respect to the size of
the statistical errors, phase switch Monte Carlo also outperformed the
thermodynamic integration method with respect to the necessary
investment of computer resources. In order to obtain a suitable weight
function for our system, at a certain value of the misfit, we let the
simulation run for about 15 million steps (each step consisting of one
sweep of local moves and one attempt to switch the phases). On the ZDV
cluster of the University of Mainz, this takes about $4.5$ days on a
single core (though in hindsight we could have got away with a less
smooth weight function, further reducing the computing time of this
step). Having determined the weight function, we initiated four
production runs for every value of the misfit. These runs needed again
10 million steps each (i.e. about 3 days each) in order to perform a
sufficient number of phase switches to yield results of the desired
precision. Overall, then, computing each point of the free energy difference
curve of Fig.~\ref{fig7} by phase switch took about $16.5$ days of CPU time.

In contrast to this, the thermodynamic integration method required a
calculation not only of the free energy difference in which we are
interested, but of the free energy difference along the path of the
thermodynamic integration, gradually switching off the wells of
attraction used there, and of the free energy difference between the
state where the particle interactions were turned on and the state where
they were turned off. This needs to be done for both phases separately.
It is therefore not surprising, that considerably more CPU time was
needed: roughly $250$ days of CPU time were invested for each phase and
for each value of the misfit to obtain the absolute free energy (again
converting units to a single core). Thus, each of the 12 values of free
energy differences needed for Fig.~\ref{fig7} required 500 days (rather
than $16.5$ days), i.e. a factor of $30$ more computational effort!
However, if we were to bring the statistical errors of the
thermodynamic integration method a factor of 10 down (to make it
comparable to the phase switch method), we would need another factor of
100 in computer time; the benefit of using the (clearly much more
powerful) phase switch approach hence amounts to a gain of the
order of 10$^3$ in computational resources! Of course, this is no
surprise when we remember that the free energy differences of interest
are only of the order of (1/500) of the total free energies for the
present model system.

\subsection{Ensemble inequivalence}

\begin{figure}
\includegraphics[scale=0.25, clip=true]{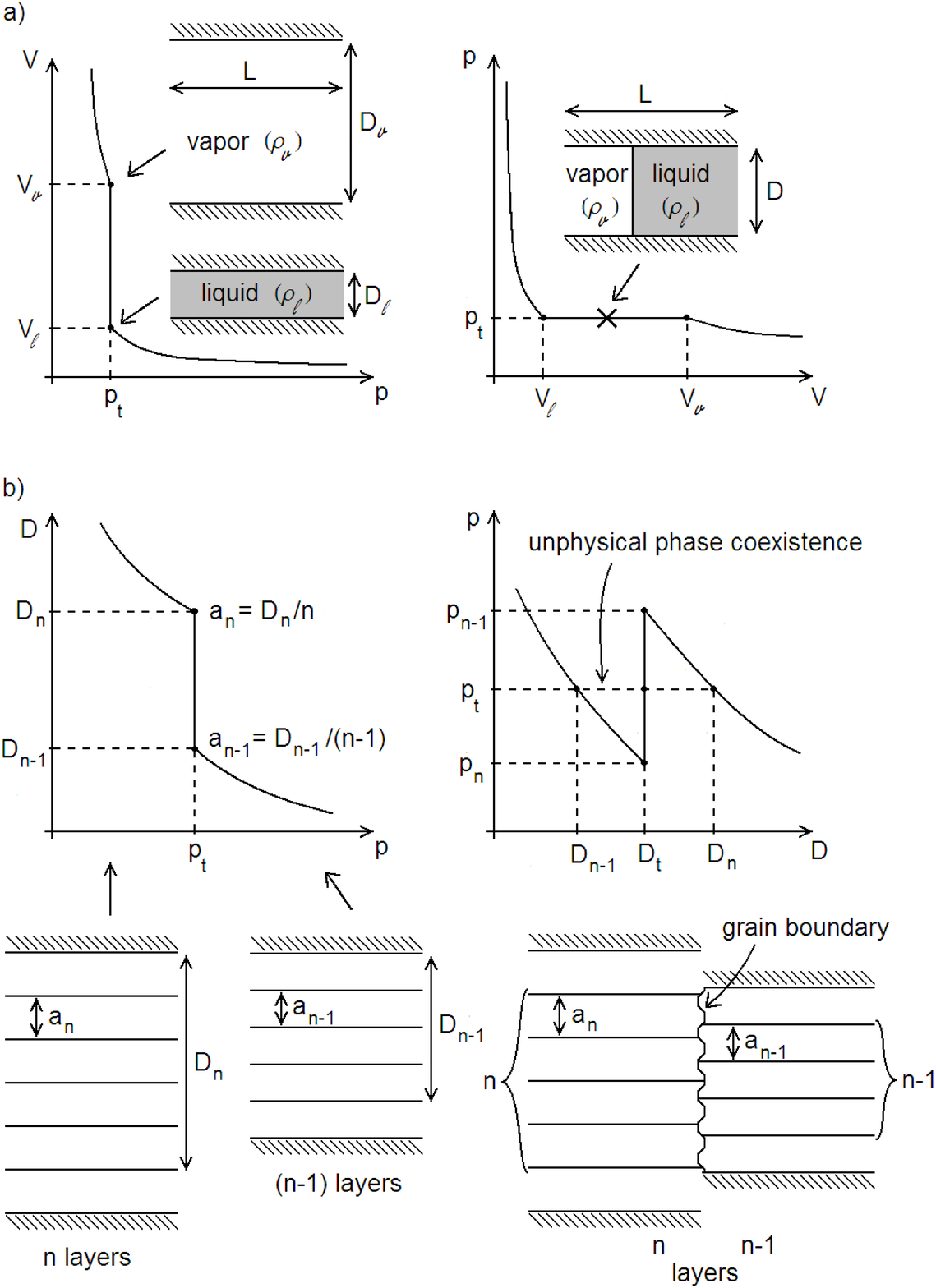}
\caption{\label{fig8} Schematic description of phase transitions in thin films of thickness $D$ in the conjugate NpT
(left) and NVT (right) ensembles, for the case of a vapor to liquid transition (a) and the present transition where the number of rows is reduced $(n \rightarrow n -1)$ when either the (normal) pressure $p$ increases (left) or the thickness decreases (right). Note that in the latter case two-phase coexistence is possible for the vapor-liquid transition, but
not for the transition where the number of rows parallel to the boundaries change. For further explanations cf text.}
\end{figure}

We turn now to a discussion of a puzzling aspect of the physics, namely
the fact that we treat here a first-order structural phase transition
obtained by variation of the distance $D$ between the walls formed by
the rigidly fixed particles, i.e. an {\it extensive} rather than an {\it
intensive} thermodynamic variable. If we were concerned with the study
of a vapor to liquid transition of a fluid in such a geometry, the
proper way to locate a discontinuous transition is the variation of the
intensive variable thermodynamically conjugate to $D$, which is the
normal pressure $p_N$ (force per area acting on the walls; in the
following the index $N$ will be omitted. Of course, at fixed lateral
dimensions $L$ a variation of $D$ is equivalent to a variation of the
volume $V$).

\begin{figure}[h!]
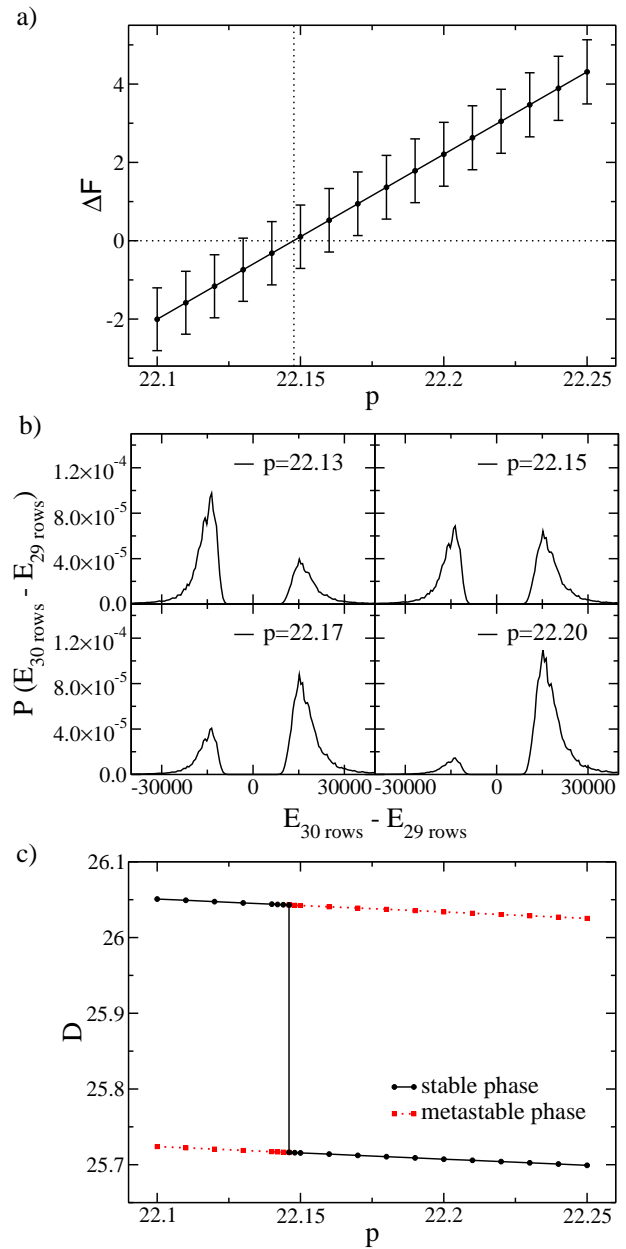

\includegraphics[scale=0.3, clip=true]{fig9a.eps}\\
\includegraphics[scale=0.3, clip=true]{fig9b.eps}\\
\includegraphics[scale=0.3, clip=true]{fig9c.eps}
\caption{\label{fig9} a) Free energy difference $\Delta F$ for the transition from $n=30$ to $n=29$ rows as a
function of pressure. (b) The distribution of the internal energy difference between the two phases $p(E_{30 rows} -E_{29 rows})$ at fixed $\{\vec{u}\}$. Curves for $4$ pressures near and at the transition pressure $p_t=22.146 \pm 0.015$ are shown, as generated via histogram reweighting. The simulation was run at a pressure of $p=22.13$. (c) System length $D$ as a function of pressure. Clearly, the curve for the stable phase exhibits a jump at the transition pressure. Statistical errors are smaller than the symbol sizes.}
\end{figure}

To fix ideas, we remind the reader about this classical vapor-liquid
problem in Fig.~\ref{fig8}a): In the NpT ensemble, we would have a jump
in volume $V=LD$ from $V_v=LD_v$ (density of the vapor $\rho_v=N/V_v)$
to $V_\ell=LD_\ell$ (density of the liquid $\rho_\ell=N/V_\ell)$ at the
transition pressure $p_t$. If we work in the conjugate NVT ensemble, of
course, the behavior simply follows from a Legendre transform, the
volume jump from $V_v$ to $V_\ell$ translates into a horizontal plateau
at $p=p_t$, and any state of this plateau is a situation of two-phase
coexistence, as schematically indicated in Fig.~\ref{fig8}a).

Of course, it is also possible to consider the present transition
between a state of $n$ rows to $n-1$ rows in the NpT ensemble
(Fig.~\ref{fig8}b and Fig.~\ref{fig9}c). Then it is clear that the transition will show up as
a jump in the thickness $D$ from $D_n(=na_n)$ to $D_{n-1}\, (=(n-1)
a_{n-1})$, where $a_n$, $a_{n-1}$ are the (average) distances between
the lattice rows (or lattice planes, in three dimensional films,
respectively). The corresponding phases of the $n$-layer state and
$(n-1)$ layer state are indicated below the isotherm in the $(p-D)$ plane
schematically.

However, one simply cannot construct a state of two-phase coexistence
out of these two ``pure phases'' at a value of $D$ intermediate between
$D_{n-1}$ and $D_n$: locally the $n$-layer state requires a thickness
$D_n$, the $(n-1)$ layer state a thickness $D_{n-1}$, so one would have
to ``break'' the walls. Of course, it is not just sufficient to have a
state with $n$ layers separated by a grain boundary from a state with
$(n-1)$ layers at the same value of $D$: these domains are {\it not} the
coexisting pure phases in the NpT ensemble!

So the phase coexistence drawn (horizontal broken curve) in
Fig.~\ref{fig8}b) is unphysical, it requires a state where the
constraining walls were broken. Requesting the integrity of the walls is
a global constraint which makes phase coexistence in the standard sense
impossible for the present transitions! Thus, the rule that the
different ensembles of statistical mechanics yield equivalent results in
the thermodynamic limit is not true for the present system; in the
transition region $D_{n-1} < D < D_n$ the NVT ensemble and the NpT
ensemble are {\it not equivalent}.

Actually this is not the first time that such an ensemble inequivalence
has been pointed out. A case much discussed in the literature is the
``escape transition'' of a single polymer chain of $N$ beads grafted at
a planar surface underneath a piston held at a distance $D$ above the
surface to compress the polymer \cite{61,62,63,64,65,66,67}. For pressures
$p<p_t$ (where the piston is at distance $D_{t,1}$) the chain is
completely confined underneath the piston (which has the cross section
of a circle in the directions parallel to the surface) while for $p >
p_t$ the chain is (partially) escaped into the region outside of where
the piston acts (the piston distance at $p_T$ jumps to a smaller value
$D_{t,2}$). When we use instead $D$ as the control variable, again a
sharp transition occurs (for $N \rightarrow \infty$) at some
intermediate value $D_t$ $(D_{t,2}< D_t <D_{t,1})$, since obviously it is
simply inconceivable to have within a single chain phase coexistence
between states ``partially escaped'' and ``fully confined'', since these
states are only defined via a global description of the whole polymer
chain.

Another case where transitions of the number $n$ of layers in layered
structures in thin films occurs is the confinement of symmetric block
copolymer melts (which may form a lamellar mesophase of period
$\lambda_0$ in the bulk) in thin films between identical walls
\cite{68,69,70,71}. When then the thickness $D$ of such films is varied,
one observes experimentally discontinuous transitions in the number $n$
of lamellae parallel to the film \cite{69,70}. However, when one
considers block copolymer films on a substrate and does not impose the
constraint of a uniform thickness but rather allows the upper surface to
be free, then indeed mixed phase configurations of a region where $n-1$
layers occur (and take a thickness $D_{n-1})$ and of a region where $n$
layers occur (and take a thickness $D_n$) are conceivable \cite{71} and
have been observed, see e.g. \cite{72}. In summary of these remarks, we
note that it is not uncommon that global geometric constraints may
destroy the possibility of phase coexistence.

In view of the above discussion, it is of interest also in the present
case to investigate the use of the (normal) pressure $p$ (instead of the
strip width $D$) as the control variable. Taking, in the spirit of the
general remarks on the phase switch method, the appropriate phase switch
energy cost as an order parameter $M$, we can sample the probability
distribution function $p(M)$ which exhibits two well separated peaks of generally different
weights. These peaks are even more clearly visible in the distribution of the energy difference $p(E_{30 rows} -E_{29 rows})$ at fixed $\{\vec{u}\}$ as the order parameter $M$ is related to this energy difference via a logarithmic function (cf. eq.~\ref{def_M}). The transition pressure $p_t$ is that for which the peaks have
equal weight (Fig.~\ref{fig9}) and can be determined accurately via histogram
reweighting. From this we estimate that $p_t=22.146
\pm 0.015$. At the transition, the measured misfit $\Delta$ jumps from
$\Delta_1=1.913 \pm 0.043$ (for $n=30$) to $\Delta_2=1.503 \pm 0.046$
(for $n=29$). Interestingly, the misfit where the transition in the NVT
ensemble occurs ($\Delta_t \approx 1.71)$ is just the average of these
two values.

\subsection{Comparison of competing candidate stable structures}

\begin{figure}[h!]
\includegraphics[scale=0.32, clip=true]{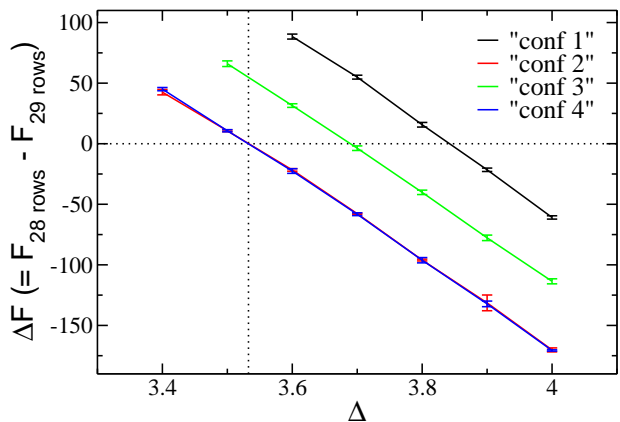}
\caption{\label{fig10} Free energy differences between various structures with $n=28$ rows and the structure with $n=29$ plotted vs. the misfit $\Delta$. As configurations nr. 2 and nr. 4 turned out to be the same, their free energy curves fall on top of each other.}
\end{figure}

Returning again to the NVT ensemble, we now consider the transition from
states with 29 layers to states with 28 layers. We recall
(Fig.~\ref{fig4}) that several different candidate structures do exist,
and it is not at all clear {\em a-priori}, which of them should be
favored. Again, the phase switch Monte Carlo is a convenient tool to
solve such a problem: we utilize reference states from all four of the
candidate structures having $n=28$ (as shown in Fig.~\ref{fig4}) and
calculate the free energy difference $\Delta F$ between the (unique)
structure with $n=29$ and these four candidates.

The results (Fig.~\ref{fig10}) clearly show that configurations number
$1$ and number $3$ are metastable, because they have distinctly higher
free energy differences throughout the range of $\Delta$ than
configurations number $2$ and $4$ which practically coincide. In fact,
this coincidence between the free energies of configurations nr. $2$ and $4$ is not accidental: a closer evaluation of
their time evolution shows that they transform into each other via
sequences of ``easy'' local moves, and although the instantaneous
snapshot pictures reproduced in Fig.~\ref{fig4} were different, they do
not belong to different phases in a thermodynamic sense.

It is also interesting to note that the conclusion that structure number
2 is the stable one would not have been obtained by a simply comparison
of the internal energies of the four structures: indeed configuration
number 2 has the highest energy of all four structures.

Thus, entropy matters in soft crystals, such as those studied here.

\section{Concluding remarks}

The principle findings of our study are two-fold: (i) We have
performed a thorough test of the suitability of the phase switch Monte
Carlo method for the task of determining the relative stability of
imperfectly ordered structures of typical soft-matter systems, where one
must deal with systems which have at least one very large linear
dimension. For such a test, it is crucial to provide full information on
the model that is studied, and to give a careful description of the
method and its implementation. Moreover we have studied precisely the
same model system by a thermodynamic integration method thereby allowing
the first like-for-like comparison between the two approaches. We find
that the results from both methods are compatible, but the accuracy that
can be achieved using phase switch MC is at least an order of magnitude
better (Fig.~\ref{fig7}), despite requiring a factor of $30$ less
computational time.  

The reasons for this efficiency gain can be appreciated from a glance at
Fig.~\ref{fig6}: the absolute free energies of our system of $3240$
particles vary from about $22000$ to $24000$ (in suitably scaled units),
for a misfit parameter $\Delta$ varying from $1$ to $2$, while the free
energy difference between the two states that we wish to compare vary
only from $-60$ to $+60$ in the same range.  These numbers illustrate
vividly the basic concept of phase switch Monte Carlo: one does better
in focusing directly on the small free energy difference between the
states that one wishes to compare, rather than extracting them
indirectly by subtracting two measurements of large absolute free
energies. Thus (in the present context at least) phase switch Monte
Carlo seems a much more powerful approach than thermodynamic
integration.  In fact, if one were to try to bring the errors of the
thermodynamic integration method down by an order of magnitude -- to make
the error bars of both methods in Fig.~\ref{fig7} comparable -- one would
have to invest a factor of 3000 more computational time. We feel that
the case of relatively small free energy differences between competing
phases and/or structures is rather typical for soft matter systems.
Indeed for many soft matter systems, such as block copolymer mesophases,
the relative magnitude of free energy differences is much less than the
factor of about $1/500$ encountered here, and hence such problems could
never be tackled successfully with thermodynamic integration methods
since the computational effort to reach the requisite accuracy would be
prohibitive.

The first problem to which phase switch Monte Carlo was applied (in the
form of the "Lattice-switch" method), evaluated the free energy
difference of perfectly ordered face-centered cubic and hexagonal close
packed crystals. Such an application  might be regarded as a somewhat
special case due to the perfect long-range order in these defect-free
crystals. However, the present work shows that the method can equally be
applied to imperfectly ordered crystals. Here, due to the confinement by
structured walls together with a misfit between the distance between the
walls and the appropriate multiple of the distance between the lattice
rows, somewhat irregular long range defect structures form along the
walls (``soliton staircase''). Additionally several similarly
ill-crystallized structures can present themselves as candidates for the
optimal structure (Fig.~\ref{fig4}). It would be absolutely impossible
to identify which is the equilibrium structure and which structures are
only metastable without the phase switch Monte Carlo method
(Fig.~\ref{fig10}).

We note that the model system that we have chosen to study
(Fig.~\ref{fig1}) could also be experimentally realized in colloidal
dispersions, though with some effort: colloids coated with polymer
brushes experience a short ranged, almost hard-sphere-like, repulsive
effective potential, and bringing them to an interface where water is on
top and air is below, rather perfect two-dimensional crystals with
triangular lattice structure form. Interference of strong laser fields
can be used to create a periodic confining potential, through which the
misfit and thus the crystal structure can be manipulated. We hope that
our study will solicit some corresponding experimental studies to show
that the proposed transitions in the number of rows in these crystalline
strips actually occur.

(ii) Our second main finding is that this type of system has an
interesting physical property, namely the inequivalence between
conjugate ensembles of statistical mechanics. When we fix the distance
$D$ between the confining ``walls'', the total particle number $N$ and
the total (two-dimensional) ``volume'' $V$ of the system, we realize the
NVT ensemble. When one studies first order transitions in the bulk using
such an ensemble containing two extensive variables ($N$, $V$), a first
order transition normally shows up as a two-phase coexistence region
(e.g., at fixed $N$ the two-phase coexistence extends from $V_I$ to
$V_{II}$). However, here such a two-phase coexistence is not possible
(Fig.~\ref{fig8}), and thus one has the unusual behaviour that at the
equilibrium in the ``constant $D$''-ensemble the conjugate intensive
variable (the normal pressure $p_N$, as well as the stress $\sigma$, cf.
Fig.~\ref{fig2}) exhibit jumps (in Fig.~\ref{fig2}, we display the
hysteresis loops, but the positions of the jumps in equilibrium can be
inferred from $\Delta F=0$ in Figs.~\ref{fig7} and \ref{fig10},
respectively). When we use a ``constant $p$''-ensemble (which is
physically reasonable if the confinement of the crystal is effected
mechanically in a Surface Force Apparatus), it is the ``volume'' (i.e.,
the distance between the walls $D$) which jumps from $D_I$ to $D_{II}$
at a well-defined transition pressure, cf. Figs.~\ref{fig8},~\ref{fig9}.

One should not confuse this ensemble inequivalence with the well-known
ensemble inequivalence between NVT and NpT ensembles in systems where
$N$ is finite: in the latter case, the ensemble inequivalence is
dominated by interfacial contributions (in the NVT-ensemble, when $V_I <
V < V_{II}$, the system is in a two-phase configuration, as suggested
for $V \rightarrow \infty$ by the ``lever rule'', but for finite $V$ the
relative contribution due to the interface between the coexisting phases
dominate the finite size effects). But for $V \rightarrow \infty$ these
interfacial effects become negligible, the properties in the two
conjugate ensembles are just related by the appropriate Legendre
transformation. This equivalence between the ensembles holds also for
liquid-vapor or liquid-liquid unmixing under confinement in a thin film
geometry: when $D$ is finite and the particle number $N \rightarrow
\infty$, i.e. the lateral linear dimensions become macroscopic, we still
have ordinary two-phase coexistence in the thin films (cf.
Fig.~\ref{fig8}). The ensemble inequivalence in the present system
arises from the lack of commensurability between the thickness $D$ of
the slit and the appropriate multiple of the lattice distance. At a
transition pressure $p_t$ in the NpT ensemble we inevitably have
different distances $D_I$, $D_{II}$ between the walls for the two phases
$I$, $II$. Thus, they cannot coexist for any uniform value of $D$.
Similar phenomena (where the number of layers of a layered lamellar
structure confined between walls exhibits jump discontinuities when $D$
is varied) are already known, both experimentally and theoretically, for
block copolymer mesophases, but the aspect of ensemble inequivalence has
not been addressed, to our knowledge, in these systems studied here.

\section{Acknowledgements}

One of us (D.W.) acknowledges support from the Deutsche
Forschungsgemeinschaft (DFG) under grant number TR6/C4 and from the
Graduate School of Excellence ``Material Science in Mainz (MAINZ)''. She
is also grateful to the Department of Physics, University of Bath
(UK), for its hospitality during an extended research stay under the
auspices of the visiting postgraduate scholar scheme. We thank P.
Virnau, T. Schilling, F. Schmid and I.M. Snook for helpful discussions
and advice.

\clearpage

\section{Appendix}
\appendix
\setcounter{figure}{0}

Here we provide an extended description of the implementation of phase
switch Monte Carlo, concentrating on implementation details at a level
suitable for a new practitioner.

\section{Implementation details for the phase switch method}
\label{append_phase_switch}

In order to calculate the free energy difference between two phases in a
single simulation run, the two phases have to be linked by a sampling path.
In many popular approaches, a direct path between the two
phases is constructed in the form of a continuous set of macrostates
associated with the values of some order parameter which distinguishes
one phase from the other (common examples are the total energy or
density of a fluid). This path traverses mixed phase (interfacial)
states \cite{74} and is negotiated using some form of extended sampling to
overcome the free energy (surface tension) barrier associated with the
interfacial states. One way to do this is the multicanonical method
\cite{60}. Alternatively one can directly measure free energy
differences between successive points along the path as is the case in
the successive umbrella sampling technique \cite{73}.

In many cases utilizing  an inter-phase path that encompasses interfacial
states works well, particularly for fluid-fluid transitions or lattice
models of magnets. However, in other cases such a path can be
problematic \cite{9}. For example in the case of solid-liquid
coexistence, a connecting path will typically run from a crystalline
phase through several different distinct states including droplets of
liquid in a crystal, a slab configuration and crystalline droplets in a
liquid before finally reaching the pure liquid phase \cite{75}. In
such cases the identification of a suitable order parameter to guide the
system smoothly from one pure phase to the other can be difficult, and
as a result the system may experience kinetic trapping (eg in defective
crystalline states).

Thus it is highly desirable to have a method which can directly ``leap''
between the two pure phases (which we shall label $\alpha$, with
$\alpha=1,2$), avoiding the problematic mixed phase states.  If the
system jumps back and forth between these phases a sufficient number of
times within one simulation run, the relative probability with which the
system is found in each of them directly yields the free energy
difference between these phases via $\Delta F=-\ln \left(
\frac{P^{(\alpha=1)}}{P^{(\alpha=2)}} \right)$. The phase switch method
achieves this by supplementing standard local particle displacement
moves (and in the case of a simulation in the NpT ensemble, moves which
scale the volume of the simulation box), with moves that switch the
system from one phase directly into the other phase. This switch is
facilitated by the {\em representation} of particle configurations in
the two phases. Specifically we associate a fixed reference
configuration $\{\vec{R}^{(\alpha)}\}$ with each phase. The reference configuration is an
arbitary configuration drawn from the set of configurations that are identifiable as
`belonging' to phase $\alpha$.  We then associate each particle with a
unique site of the reference configuration, allowing us to write its
position $\vec{r}_i^{(\alpha)}$ in terms of the displacement $\vec{u}_i$
from its reference site:

\begin{equation}
\vec{r}_i^{\,(\alpha)}= \vec{R}_i^{\,(\alpha)} + \vec{u}_i
\end{equation}

Note that whilst there are two reference configurations (one for each
phase), the phase switch method only considers one set of displacement vectors which are regarded
as common to both phases.

Suppose we are currently in phase $\alpha=1$, so that the  particle
coordinates are $\vec{r}_i^{\,(1)}= \vec{R}_i^{\,(1)} + \vec{u}_i$. For
local moves in this phase we update particle coordinates (in the manner
to be described) which, owing to reference sites being fixed, is
equivalent to updating the displacement vectors. For a phase switch to
phase $\alpha=2$, we propose a new configuration which is simply formed
by substituting the reference sites of phase $\alpha=1$ with those of
phase $\alpha=2$. Thus the proposed
configuration is $\{\vec{r}_i^{\,(2)}\}= \{\vec{R}_i^{\,(2)}\} +
\{\vec{u}_i\}$. If this switch is accepted, i.e. if the resulting
configuration of phase $\alpha=2$ is energetically acceptable, the
simulation will continue to run in phase $\alpha=2$, again recording the
displacements of all of the particles from the reference sites of phase
$\alpha=2$, and proposing switches back to phase $\alpha=1$.  In this
way the system switches repeatedly back and forth between the phases,
allowing one to record the relative probability of finding the system in
each phase.

Now generally speaking the displacement vectors that characterise phase
$\alpha=1$ are not typical of phase $\alpha=2$ and thus it will not be
energetically acceptable to perform the switch from typical
configurations of phase $\alpha=1$. To deal with this, one introduces a
bias in the accept/reject probabilities for local moves that enhances
the probability of displacements being generated in phase $\alpha=1$ for
which the phase switch to $\alpha=2$ {\em is} energetically acceptable.
The obvious observable to which the bias should be administered is a
quantity related to the instantaneous energy cost of the switch, since
this measures how likely it is to be accepted. We have employed the
switch energy order parameter $M$ described in Ref.~\cite{21}, which for
switches from phase $\alpha=1$ to $\alpha=2$ is defined as follows:

\begin{equation}
M^{(1)\to(2)}(\{ \vec{u} \}) = \rm{sgn}(\Delta E^{(1)\to(2)}) \cdot \ln(1+|\Delta E^{(1)\to(2)}|)
\label{def_M}
\end{equation}

where

\begin{equation}
\Delta E^{(1)\to(2)} = (E^{(2)}(\{\vec{u}\}) - E^{(2)}_{ref})-(E^{(1)}(\{\vec{u}\}) - E^{(1)}_{ref})
\label{def_M_2}
\end{equation}
where $E^{(\alpha)}_{ref}$ is the energy of the reference configuration in
phase $\alpha$, and $E^{(\alpha)}(\{\vec{u}\})$ is the energy in phase $\alpha$, found by applying the displacement vectors $\{\vec{u}\}$ to the reference configuration $\{\vec{R}^{(\alpha)}\}$. An obvious substitution gives the order parameter for
the switch from $\alpha=2$ to $\alpha=1$. Note that an important feature of this definition of this order parameter is
the logarithm which ensures that the binning of the weight function is
finer for small values of the energy difference and thus serves to
ensure that the simulation can cover the entire range of $M$ smoothly. 

Now, when implementing local moves for particles, we consider not just the energy cost of the
move within the current phase, but also the change in $M$ associated with the local
move via a weight function $\eta(M)$. The acceptance criterion for the local move is therefore given by:

\begin{align}
&p^{(\alpha)}(\{ \vec{u}\} \rightarrow \{ \vec{u'} \}) =\nonumber\\ &\rm {min}(1, e^{-\beta (E^{(\alpha)}(\{ \vec{u'} \}) - E^{(\alpha)}(\{ \vec{u} \})) + \eta(M')-\eta(M)}).
\label{metrop_local_phase_switch}
\end{align}
Note that $E^{(\alpha)}(\{
\vec{u'} \})) - E^{(\alpha}(\{ \vec{u} \})$ is the energy difference
due to the move in the phase $\alpha$ that is currently being simulated. The
energy difference in the other phase is only needed for the computation
of the new order parameter $M'$ and therefore for the weights $\eta(M')$ associated
with the move. 

Phase switches are generally only accepted from states in which $M$
is small -- the so called gateway states. One instance in which 
$M$ becomes small is if the displacement vectors are themselves small,
i.e. if all particles are sitting close to their reference positions in
both phases. Another instance is if there is a high degree of structural
similarity among the phases, so that the displacements of many of the
particles in one phase are typical of the displacements in the other
phase. Note that one does not need to know or specify the gateway states
to use the method. They are sought out automatically when one
biases to small values of $M$.

The acceptance criterion for a phase switch from $\alpha=1$ to $\alpha=2$ itself reads:

\begin{equation}
p^{(1)\to(2)}(\{ \vec{u} \})={\rm min}(1, e^{-\beta (E^{(2)}(\{ \vec{u} \}) - E^{(1)}(\{ \vec{u} \}) + \omega^{(2)}-\omega^{(1)}})\:,
\label{metrop_PS_phase_switch}
\end{equation}
and similarly for the reverse switch. This phase switch also includes a weight $\omega$ to ensure that it occurs
with a sufficiently high probability in both directions. Note that since
the phase switch move alters the absolute particle coordinates, the
associated energy change enters the switch acceptance criterion. In the
case of phase switch simulations in the NpT ensemble, an additional
volume scaling, must also be taken into account, see below.

\begin{figure}[htbp]
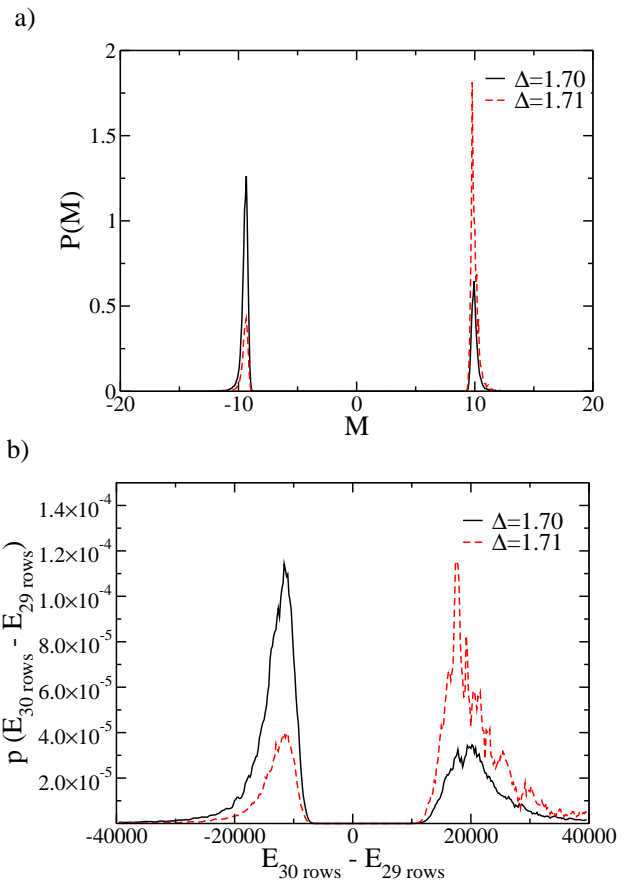

\centering
\includegraphics[scale=0.3, clip=true]{fig_A1a.eps}\\
\includegraphics[scale=0.3, clip=true]{fig_A1b.eps}
\caption[KURZFASSUNG]{a) The order parameter distribution $p(M)$ for simulations at $\Delta=1.70$ and $\Delta=1.71$ carried out in the NVT ensemble. b) For comparison the same distribution is plotted against the internal energy difference between the two phases for fixed $\{\vec{u}\}$. The order parameter $M$ is deduced from this energy difference $E_{29 rows} -E_{30 rows}$ via the definition given in eq.~\ref{def_M}.}
\label{Graph_peaks_NVT}
\end{figure}

Once suitable weights have been determined (see appendix \ref{append_transfer_matrix}), one samples the statistics
of the two phases by accumulating a histogram of the biased order parameter
distribution $\tilde P(M)$. At the end of the simulation, the effects of
the weights is unfolded from this distribution in the standard manner
for extended sampling \cite{9} to find the equilibrium distribution
$P(M)$. Close to a phase transition, this distribution will exhibit two
well separated peaks, whose areas yield the free energy difference as
described above. An example is shown in Fig.~\ref{Graph_peaks_NVT} (a). Also show in Fig.~\ref{Graph_peaks_NVT} (b) is the distribution of the instantaneous energy change under the switch $E^{(\alpha^\prime)}(\{\vec{u}\})-E^{(\alpha)}(\{\vec{u}\})$ which similarly shows two peaks, one for each phase.

With regard to the choice of reference configuration in each phase, in
principle this can be an arbitrary configuration belonging to that
phase. In practice, however, for crystalline systems one finds that the degree of
weighting required to access the gateway states can be reduced by
choosing a reference configuration which is a perfect lattice. For more
general system, eg those with crystalline disorder, or for fluids it may
be advantageous to try to ensure that the particles are not sitting too
close to each other (eg. by energy minimization of the configuration
\cite{20}), since particles which are in close proximity reduce the
number of gateway states significantly. (We note in passing that for
fluid systems \cite{21} one requires special approaches to guide
particles to the gateway states that we will not discuss here as they
were not necessary for our system.)

\begin{figure}[htbp]
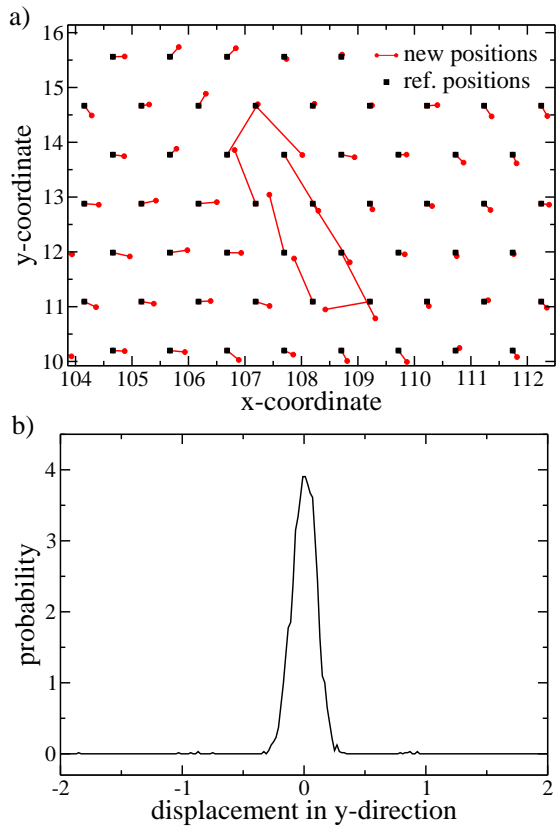

\centering
\includegraphics[scale=0.3, clip=true]{fig_A2a.eps}
\includegraphics[scale=0.3,clip=true]{fig_A2b.eps}
\caption[KURZFASSUNG]{a) A section of a configuration that includes
particles which have swapped positions in the phase with 29 rows. Black
squares denote the reference configuration, red (grey) lines are the
displacement vectors associated with the reference positions and red
(grey) dots are the new positions. This
simulation was carried out with 3240 particles at a misfit of
$\Delta=1.70$ in an NVT simulation with 1280 bins for the weight
function. b) Typical histograms of the particle displacements
in y-direction. The (almost completely invisible) very small peaks at about $+1$ and $-1$ correspond to particles
which have swapped their positions.}
\label{Graph_stuck_in_swap}
\end{figure}

With regard to the phase switch simulations of the present model of 2d
colloids in confinement, we mention a rare problem that appeared in our
simulations of the $29$ row system. This involved sets of particles on
neighbouring lattice sites in adjacent rows jumping between rows during
the simulation,  creating in the process a ring of particles which
occupy each others positions (cf. Fig.~\ref{Graph_stuck_in_swap} (a)) and
remain there. This occurrence is primarily a feature of the
two-dimensional nature of our system, and the well known 'softness' of
2d crystals. When it occurs it interferes with the operation of the
phase switch method because the weight function is not designed to deal
with it, so one is prevented from reaching the gateway states. Although
one can envisage methods for solving this problem along the lines of
those used in fluids \cite{21}, our solution to the problem was to
simply suppress it. Measurements of the distribution of displacements in
the y-direction is shown in Fig.~\ref{Graph_stuck_in_swap} (b) and
show that preventing particles from fluctuating any further in in
y-direction than $\Delta y=0.5$ introduces a negligible constraint with
regard to their natural fluctuations (and hence on free energy
measurements). Doing so cured the problem of rare lattice site swaps.

Finally, we outline briefly how to apply the phase
switch method in the NpT ensemble. The advantage of the NpT ensemble is
that the results obtained at one pressure can easily be extrapolated to
other values of the pressure by standard histogram reweighting methods.
The difference between the NVT and the NpT ensemble in this case is that
additional volume moves have to be carried out in the NpT ensemble. In
such moves, all particle coordinates are scaled (along with the box) in
both phases in the standard way \cite{6,21}. Additionally, it can prove
useful to combine the phase switch move itself with a volume scaling
move if the equilibrium densities of the two phases differ from each
other as it was the case for our system. For details on the underlying
statistical dynamics, and acceptance probabilities, see Ref. \cite{21}.
The problem of particles switching their positions and thus creating configurations which prevented any further phase switches from being accepted did not occur in the case of simulations in the NpT ensemble for our system. We obtained (within the error bars) the same free energies whether or not we restricted the movement of the particles in the $y$-direction in the way we had to restrict them in the NVT ensemble.

\setcounter{figure}{0}
\section{Implementation details for the transfer matrix method}
\label{append_transfer_matrix}

The choice of method for determining the weight function $\eta(M)$ that
connects the configurations of high statistical weight to the gateway
states is to some extent a matter of personal taste. A number of
approaches exist such as the Wang-Landau method \cite{59} or successive umbrella
sampling \cite{73}. In this work, we have found the transition matrix method to be
a particularly efficient means of determining a suitable weight
function. The transition matrix method has the advantage that - similar
to the Wang-Landau-Sampling - the weights can be updated ``on the fly''
throughout the simulation, allowing the simulation to  explore an ever
wider range of values of the order parameter $M$ as the weight function
evolves, until it eventually encompasses the gateway states of low $M$.
Once this has been achieved, one can cease updating the weight function
and perform a simulation run with a constant weight function. An
advantage of transition matrix method over Wang-Landau sampling is that
it collects equilibrium data from the outset of the simulation, whereas
Wang-Landau only provides equilibrium estimates after a number of
preliminary iterations.

The general idea of the transition matrix method for determining weight functions is to
record the acceptance probabilities of all attempted transitions and
extract the ratio of the states' probabilities from it. As all attempted
transitions contribute to the weight function, including those that were
rejected, the weight function can be built up rather quickly. The
details of the implementation are as follows and can also be found in
\cite{20,21,56a} and the references given therein.

To implement the transition matrix method, the range of the order
parameter $M$, for which a weight function is desired, is divided into a
number of bins. In our case this range corresponds to the values of $M$
that lie between the peaks in $P(M)$ which correspond to the two phases (cf. Fig.~\ref{Graph_peaks_NVT} (a). A good choice for the binning of the order parameter is to choose the
bins in such a way, that the weight difference between adjacent bins
satisfies \cite{21} $|\eta(M_{i+1})-\eta(M_i)| < 2$.
Then, for every attempted move the acceptance probability $p$, (which is
calculated anyway for use in  the Metropolis criterion) is stored in a
collection matrix $C$:

\begin{equation}
C(M\rightarrow M') \Rightarrow C(M\rightarrow M') + p
\label{C_1}
\end{equation}
At the same time, the probability for rejecting the move and thereby keeping the current value of the order parameter is also stored:
\begin{equation}
C(M\rightarrow M) \Rightarrow C(M\rightarrow M) + (1-p)
\label{C_2}
\end{equation}
It is important to note that these probabilities $p$ are the ``bare'' acceptance probabilities and do not include any weights.

The transition probabilities are then simply calculated by a
normalization of the values in the collection matrix, with the sum on
the right hand side including all possible states to which the system
can jump from a given state:

\begin{equation}
T(M\rightarrow M') = \frac{C(M\rightarrow M')}{\sum_k C(M\rightarrow M_k)}
\label{C_3}
\end{equation}

In the most general case, this method would create an $N \times N$
matrix, $N$ being the number of bins or values of the order parameter
$M$. In order to derive the correct probability distribution from such an $N
\times N$ transition matrix, it is necessary to compute the eigenvector
to the largest eigenvalue of this matrix. However, it is not necessarily
required to know the exact probability distribution in order to create a
weight function that will work sufficiently well. Therefore it is
possible to take only those transitions occuring between neighbouring
bins of the order parameter into account when computing the weight
function. In terms of the transition matrix, this means that only the
diagonal elements - corresponding to transitions from a state to itself
- and the first off-diagonal elements - corresponding to transitions
from one state to the adjacent ones - are taken into account. 
Using this approach the weight function can be calculated
quite easily without the need to compute eigenvalues or eigenvectors
of the transition matrix. In this case, the ratio of the probabilities
of two adjacent states can be read off directly from the transition
matrix via

\begin{equation}
\frac{P(M_{i+1})}{P(M_i)} = \frac{T(M_i \rightarrow M_{i+1})}{T(M_{i+1} \rightarrow M_i)}
\label{trans_simplified_1}
\end{equation}
yielding the weight difference
\begin{align}
&\eta(M_{i+1})-\eta(M_i) = -\ln\left(\frac{P(M_{i+1})}{P(M_i)}\right) = \nonumber\\&-\ln\left(\frac{T(M_i \rightarrow M_{i+1})}{T(M_{i+1} \rightarrow M_i)}\right).
\label{trans_simplified_2}
\end{align}

Of course, when running the simulation, the system is still free to
perform transitions between any values of $M$. But these transitions
are not registered in the transition matrix and thus are also not taken
into account when calculating the weights. In the present study this was
found to produce accurate and useful weight functions  as transitions
between distant values of $M$  were rare and the entries in the second
off-diagonal elements of the transition matrix were already considerably
smaller than the ones we used for the calculation of the weights.

\begin{figure}[htbp]
\centering
\includegraphics[scale=0.3, clip=true]{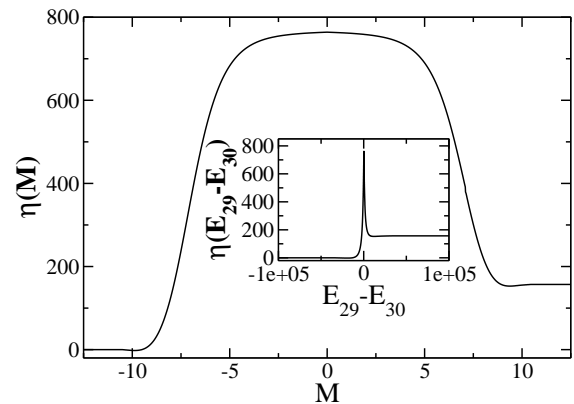}

\caption{Weight functions for the two-dimensional colloidal crystal with
$N=3240$ particles and structured walls at a misfit of $\Delta=1.7$. The
left minimum corresponds to states where the system was simulating the
phase with 29 rows, the right minimum corresponds to 30 rows. Note that
the weights have an exponential influence on the acceptance criterion.
The large figure shows the weights plotted against the order parameter
$M$ as defined in eq. \ref{def_M}, the inset shows the same weight
function plotted against the energy difference between the two phases in
order to illustrate how the definition of the order parameter in the
logarithm of the energy difference stretches the part around $M=0$,
where phase switches are most likely to happen.}
\label{Graph_weight_functions}
\end{figure}

By accumulating the transition matrix in the course of a simulation, one
obtains an estimate for $P(M)$ which can be used to update $\eta(M)$,
thereby allowing the simulation to explore a wide range of $M$.
Repeated updates of $\eta(M)$ thus extend systematically the range of $M$ over which
one accumulates statistics for the weight function, until ultimately one
reaches the gateway states. However since updating the weight function during a simulation violates
detailed balance, we chose to do this at rather infrequent intervals of
$20000$ sweeps. Once the weight function extends to the gateway states,
we stop updating the transition matrix and perform a long phase switch simulation
with a fixed weight function in order to accumulate equilibrium free
energy data.

An example of a weight function created for the system with $N=3240$
particles (plus $432$ fixed wall particles) at a misfit of $\Delta=1.7$
is given in fig. \ref{Graph_weight_functions}, also illustrating how the
definition of the energy order parameter $M$ given in eq. \ref{def_M},
which includes a logarithm of the energy difference, leads to a finer
binning in the part closer to $M=0$, where the phase switches are most
likely to happen. In fact to ensure that the transition matrix estimate
of the weight function was sufficiently smooth and reliable in this
region we reduced the number of bins somewhat.


\begin{thebibliography}{0}
\bibitem{53} Note that in \cite{23,24,25,26} the walls were oriented along the $y$ rather than the $x$-direction
\end{thebibliography}
 \end{document}